\documentclass[12pt]{article}
\begin{document}

\centerline{\Large\bf PQCD approach to exclusive $B$ decays\footnote{
based on talks presented at Flavor Physics and CP Violation, Philadelphia,
USA, May 2002, at the 3rd Workshop on Higher Luminosity B Factories,
Kanagawa, Japan, Aug. 2002, and at Summer Institute 2002, Yamanashi,
Japan, Aug. 2002}}

\vskip 1.0cm

\centerline{Hsiang-nan Li}
\vskip 0.3cm
\centerline{Institute of Physics, Academia Sinica,
Taipei, Taiwan 115, Republic of China}

\vskip 1.0cm
\centerline{Abstract}

I review the recent progress on the perturbative QCD approach to
exclusive $B$ meson decays, discussing the comparison of collinear
and $k_T$ factorizations, the CP asymmetry in the $B^0\to\pi^+\pi^-$ 
decay, penguin enhancement, the branching ratio of the 
$B^0\to D^0\pi^0$ decay, and three-body nonleptonic decays.

\section{Introduction}

Both collinear and $k_T$ factorizations are the fundamental tools of
perturbative QCD (PQCD), where $k_T$ denotes parton transverse momenta.
For inclusive processes, such as deeply inelastic scattering (DIS) of
a hadron, collinear ($k_T$) factorization applies when the process is
measured at a large (small) Bjorken variable $x_B$ \cite{Ste}
(\cite{CCH,CE,LRS}). For exclusive processes, collinear factorization was
developed in \cite{BL,ER,CZS,CZ}, in which a physical quantity is
written as the convolution of a hard amplitude with hadron distribution
amplitudes in parton momentum fractions $x$. In the inclusive
case the range of a momentum fraction $x\ge x_B$ is experimentally
controllable. In the exclusive case the range of a momentum 
fraction is not controllable, and must be integrated over between 0 and 1.
Hence, one must deal with the end-point region with a small $x$. If a
hard amplitude does not develop an end-point singularity, collinear 
factorization works. If such a singularity occurs, collinear factorization
breaks down, and $k_T$ factorization \cite{BS,LS} should be employed.

Based on the concepts of collinear and $k_T$ factorizations, the PQCD  
\cite{KLS} and QCD factorization (QCDF) \cite{BBNS} approaches
to exclusive $B$ meson decays have been developed, respectively. 
The soft-collinear effective theory is a systematic framework for the
study of collinear factorization \cite{CWB}. 
As applying collinear factorization to the semileptonic decay
$B\to \pi \ell{\bar \nu}$ at large recoil, an end-point singularity from 
$x\to 0$ was observed \cite{SHB}. According to the above explanation, we 
conclude that exclusive $B$ meson decays demand $k_T$ factorization
\cite{LY1,Kr,Me}. As shown in \cite{NL}, predictions for exclusive
processes derived from $k_T$ factorization are gauge-invariant.
In the PQCD approach the $B\to h_1h_2$ decay amplitude is written as the
convolution \cite{CL,YL,CLY,L365},
\begin{eqnarray}
A=\phi_B\otimes H^{(6)}\otimes \phi_{h 1}\otimes \phi_{h 2}\otimes
S\;,
\label{six}
\end{eqnarray}
where the six-quark amplitude $H^{(6)}$ corresponds to the diagrams with
a hard gluon emitted from the spectator quark \cite{WYL}, and 
$S$ denotes the Sudakov factor. 

Once parton transverse momenta $k_T$ are included \cite{LY1}, the
end-point singularities from small momentum fractions in exclusive $B$
meson decays are smeared. The resummation of the resultant double
logarithms $\alpha_s\ln^2(Pb)$, where $P$ denotes the dominant light-cone
component of a meson momentum, and $b$ is the variable conjugate to $k_T$,
leads to a Sudakov form factor $\exp[-s(P,b)]$. This factor suppresses the
long-distance contributions from the large $b$ region with
$b\sim 1/\bar\Lambda$, where $\bar\Lambda\equiv M_B-m_b$, $M_B$ ($m_b$)
being the $B$ meson ($b$ quark) mass, represents a soft scale. The suppression renders $k_T^2$ flowing into the hard amplitudes of 
$O(\bar\Lambda M_B)$. The off-shellness of internal particles then remain
of $O(\bar\Lambda M_B)$ even in the end-point region, and the 
singularities are removed. Since the end-point singularities do not exist 
\cite{LY1,TLS,L5,WY}, the arbitrary cutoffs introduced in QCDF 
\cite{BBNS,BBNS2} are not necessary. Therefore, factorizable,
nonfactorizable and annihilation amplitudes can be estimated in a more
consistent way in PQCD than in QCDF.

\section{CP Asymmetries}

Phenomenological consequences for two-body nonleptonic $B$ meson 
decays derived from collinear and $k_T$ factorizations are quite different.
Here we mention only the predictions for the CP asymmetry in the
$B_d^0\to\pi^+\pi^-$ decay. According to the QCDF power counting rules
\cite{BBNS,BBNS2} based on collinear factorization, the factorizable
emission diagram gives the leading contribution of $O(\alpha_s^0)$, since
the $B\to\pi$ form factor is not calculable. Because
the leading contribution is real, the strong phase arises from the 
factorizable annihilation diagram, being of $O(\alpha_s m_0/M_B)$, and 
from the vertex correction to the leading diagram, being of $O(\alpha_s)$. 
For $m_0/M_B$ slightly smaller than unity, the vertex correction 
is the leading source of strong phases. In $k_T$ factorization the
power counting rules change \cite{CKL}. The factorizable emission diagram
is calculable and of $O(\alpha_s)$. The factorizable annihilation diagram
has the same power counting as in QCDF. The vertex correction becomes of
$O(\alpha_s^2)$. Therefore, the annihilation diagram contributes the
leading strong phase. This is the reason the strong phases derived from
PQCD and from QCDF are opposite in sign, and the former has a large 
magnitude. The detailed reason is referred to \cite{CKL}.
As a consequence of the different power counting rules, QCDF 
prefers a small and positive CP asymmetry $C_{\pi\pi}$ \cite{Ben}, while
PQCD prefers a large and negative $C_{\pi\pi}\sim -30\%$
\cite{KL,LUY,LMY,US,Keum}. In the near future the two approaches to
exclusive $B$ meson decays, collinear and $k_T$ factorizations, could be 
distinguished by experiments \cite{Nir,Ros}.
  
Significant CP asymmetries are also expected in the 
$B\to K\pi$ \cite{KLS}, $B\to KK$
\cite{CHL} and $B\to \rho K$, $\omega K$ \cite{CHC} decays. The last two
modes are especially sensititve to the annihilation contributions. It has
been pointed out \cite{BG} that contribution from intrinsic charms, one of
the higher $B$ meson Fock states, reduces the magnitude but does not flip
the sign of the CP asymmetries in the $B\to K\pi$ decays. It implies that
PQCD has caught the correct leading picture of exclusive $B$ meson decays.

\section{Penguin Enhancement}

The leading factorizable contributions involve four-quark hard
amplitudes in QCDF, but six-quark hard amplitudes in PQCD. This
distinction also implies different characteristic scales in the two
approaches: the former is characterized by $m_b$, while the latter is
characterized by the virtuality of internal particles of order
$\sqrt{\bar\Lambda M_B}\sim 1.5$ GeV \cite{KLS,KL,LUY}. It has been known
that to accommodate the $B\to K\pi$ and $\pi\pi$ data, penguin
contributions must be large enough. In QCDF one relies on the chiral
enhancement by increasng the chiral symmetry breaking scale to a large
value $m_0\sim 3$-4 GeV \cite{WS}. Because of the renormalization-group
evolution effect of the Wilson coefficients, the lower hard scale leads 
to the dynamical penguin enhancement in PQCD.

Whether the dynamical enhancement or the chiral enhancement
is responsible for the large $B\to K\pi$ branching ratios can be tested
by measuring the $B\to \phi K$ modes \cite{CKL,L6}. In these modes
penguin contributions dominate, such that their branching ratios are
insensitive to the variation of the unitarity angle $\phi_3$. Because
the $\phi$ meson is a vector meson, the mass $m_0$ is replaced by the
physical mass $M_\phi\sim 1$ GeV, and the chiral enhancement does not
exist. If the branching ratios of the $B\to\phi K$ decays are around
$4\times 10^{-6}$ \cite{HMW,CY}, the chiral enhancement may be essential
for the penguin-dominated decay modes. If the branching ratios
are around $10\times 10^{-6}$ as predicted in PQCD \cite{CKL,M},
the dynamical enhancement may be essential. However, it should be 
mentioned that the infrared cutoffs in QCDF have been assumed to be 
different in the $B\to PP$ and $VP$ decay amplitudes \cite{DZ}. 
Introducing two independent sets of free parameters for the $B\to PP$ and
$VP$ modes, the $B\to\phi K$ branching ratios can be fit without 
increasing the $B\to K\pi$ branching ratios. Nevertheless, the 
$B\to\phi K$ branching ratios are enhanced by large annihilation
contributions \cite{DZ}, which seem to violate the QCDF power counting
rules.

\section{$B\to D^0\pi^0$}

The PQCD formalism for $B\to D^{(*)}$ transitions has been developed
\cite{TLS2}, which holds under the hierachy,
\begin{eqnarray}
M_B\gg M_{D^{(*)}}\gg \bar\Lambda\;,
\label{ll}
\end{eqnarray}
with $M_{D^{(*)}}$ being the $D^{(*)}$ meson mass. The relation
$M_B\gg M_{D^{(*)}}$ justifies perturbative evaluation of the
$B\to D^{(*)}$ form factors at large recoil and the definition of
light-cone $D^{(*)}$ meson wave functions. The relation  
$M_{D^{(*)}}\gg\bar\Lambda$ justifies the power expansion in the 
parameter $\bar\Lambda/M_{D^{(*)}}$. We have calculated the
$B\to D^{(*)}$ form factors as double expansions in $M_{D^{(*)}}/M_B$
and in $\bar\Lambda/M_{D^{(*)}}$, and found that the leading PQCD
factorization formulas respect the heavy-quark symmetry.

Based on the power counting rules constructed in \cite{TLS2},
it can be shown that the relative importance of the different topologies
of diagrams for the $B\to D\pi$ decays is given by
\begin{eqnarray}
{\rm emission} : {\rm nonfactorizable} 
=1 : \frac{M_D}{M_B}\;,
\end{eqnarray}
which approaches $1:\bar\Lambda/M_B$ as the $D$ meson mass $M_D$ reduces 
to the pion mass of $O(\bar\Lambda)$. Since the factorizable and
nonfactorizable diagrams contribute to the parameters $a_1$ and $a_2$ in
PQCD, respectively, the ratio $|a_2|/a_1\sim 0.5$ is obtained.
Moreover, the imaginary nonfactorizable amplitudes determine the
relative phase of the factorizable and nonfactorizable contributions,
which is about $-57^o$. The PQCD predictions
for the $B\to D\pi$ branching ratios \cite{KLL},
\begin{eqnarray}
& &B(B^-\to D^{0}\pi^-)\sim 5.5\times 10^{-3}\;,
\nonumber\\
& &B({\bar B}^0\to D^{+}\pi^-)\sim 2.8\times 10^{-3}\;,
\nonumber\\
& &B({\bar B}^0\to D^{0}\pi^0)\sim 2.6\times 10^{-4}\;,
\end{eqnarray}
are consistent with the experimental data, including those recently
observed for the ${\bar B}_d\to D^{(*)0}\pi^0$ decay
\cite{Belle,CLEO,Bab}. Hence, we are not convinced by the conclusion 
drawn from the factorization assumption that the
$B\to D^0\pi^0$ data hint large final-state interaction 
\cite{NPe,X,C,CHY} (see also \cite{JPL}).

\section{Three-body Decays}

Three-body nonleptonic $B$ meson decays have been observed recently
\cite{Bel,Bar}. Viewing the experimental progress, it is urgent to
construct a reliable framework for these modes. Motivated by its
theoretical self-consistency and phenomenological success, we have
generalized PQCD to three-body nonleptonic $B$ meson decays \cite{CL2}.
A direct evaluation of the hard amplitudes, which contain two virtual
gluons at lowest order, is, on one hand, not practical due to the enormous
number of diagrams. On the other hand, the region with the two gluons
being hard simultaneously is power-suppressed and not important.
Therefore, a new input is necessary in order to catch dominant
contributions to three-body decays in a simple manner.
The idea is to introduce two-meson distribution amplitudes \cite{MP}, by
means of which the analysis is simplified into the
one for two-body decays (number of diagrams is greatly
reduced). Both nonresonant contributions and resonant contributions through
two-body channels can be included. Moreover, the application
of this formalism to three-body baryonic decays \cite{CCY}
is straightforward.

Here we pick up the leading term in the complete Gegenbauer expansion of
the two-pion distribution amplitudes $\Phi(z,\zeta,w^2)$ \cite{MP}:
\begin{eqnarray}
\Phi_{v}(z,\zeta,w^2)&=&\frac{3F_{\pi}(w^2)}{\sqrt{2N_c}}z(1-z)(2\zeta-1)\;,
\nonumber\\
\Phi_{s,t}(z,\zeta,w^2)&=&\frac{3F_{s,t}(w^2)}{\sqrt{2N_c}}z(1-z)\;,
\label{2pi}
\end{eqnarray}
where the subscripts $v$, $s$ and $t$ stand for the vector, scalar and
tensor components, respectively. The variable $z$ ($\zeta$) is the parton
(pion) momentum fraction, and $F_{\pi,s,t}(w^2)$ the time-like pion
electromagnetic, scalar and tensor form factors with $F_{\pi,s,t}(0)=1$.
That is, the two-pion distribution amplitudes are normalized to the
time-like form factors.

To calculate the nonresonant contribution, we propose the parametrization,
\begin{eqnarray}
F^{(nr)}_\pi(w^2)=\frac{m^2}{w^2+m^2}\;,\;\;\;\;
F^{(nr)}_{s,t}(w^2)=\frac{m_0 m^2}{w^3+m_0 m^2}\;,
\label{non}
\end{eqnarray}
where the the parameter $m=1$ GeV is determined by the fit to
the  experimental data $M_{J/\psi}^2|F_\pi(M_{J/\psi}^2)|^2\sim 0.9$
GeV$^2$ \cite{PDG}, $M_{J/\psi}$ being the $J/\psi$ meson mass. 
To calculate the resonant contribution, we parametrize
it into the time-like form factors,
\begin{eqnarray}
F^{(r)}_{\pi,s,t}(w^2)=\frac{M_V^2}{\sqrt{(w^2-M_V^2)^2+\Gamma_V^2w^2}}
-\frac{M_V^2}{w^2+M_V^2}\;,
\label{res}
\end{eqnarray}
with $M_V$ ($\Gamma_V$) being the mass (width) of the resonance meson 
$V$. The subtraction term renders Eq.~(\ref{res}) exhibit
the features of resonant contributions: the normalization
$F^{(r)}_\pi(0)=0$ and the asymptotic behavior
$F^{(r)}_\pi(w^2)\sim 1/w^4$, which decreases at large $w$ faster than
the nonresonant parametrization in Eq.~(\ref{non}). Equation~(\ref{res})
is motivated by the pion time-like form factor measured at the $\rho$
resonance \cite{RR}. 

For the $B^+\to\rho^0(770) K^+$
and $B^+\to f_0(980) K^+$ channels, we choose $\Gamma_\rho=150$ MeV
and $\Gamma_{f_0}=50$ MeV \cite{KH}.
The nonresonant contribution $0.61\times 10^{-6}$ to the $B^+\to
K^+\pi^+\pi^-$ branching ratio is obtained. Our results $1.8\times
10^{-6}$ and $13.2\times 10^{-6}$ are consistent with the measured
three-body decay branching ratios through the $B^+\to \rho(770)
K^+$  and $B^+\to f_0(980) K^+$ channels, $< 12\times 10^{-6}$ and
$(9.6^{+2.5+1.5+3.4}_{-2.3-1.5-0.8})\times 10^{-6}$ \cite{Bel},
respectively. Since the $f_0$ width has a large uncertainty, we
also consider $\Gamma_{f_0}=60$ MeV, and the branching ratio
reduces to $10.5\times 10^{-6}$. The resonant contributions from
the other channels can be analyzed in a similar way. For example,
the $K^*(892)$ resonance can be included into the $K$-$\pi$ form
factors by choosing the width $\Gamma_{K^*}=50$ MeV. 

\section{Conclusion}

I have briefly reviewed the PQCD approach to two-body nonleptonic $B$
meson decays. This approach is based on the rigorous $k_T$ factorization
theorem, in which both meson wave functions and hard amplitudes can be
constructed in a gauge-invariant way. The predictions for the branching
ratios and the CP asymmetries of varous modes are in agreement with the
experimental data \cite{Keum}. The $B\to K\eta'$ data are an exception,
which may indicate a significant gluon content of the $\eta'$ meson
\cite{KS,AP,BN}. The charm penguin contribution could be significant, and
requires more investigation \cite{S,MS}. The generalization to three-body
nonleptonic decays seem to be successful. In the future we shall
work out the next-to-leading-order and next-to-leading-power
corrections, which are expected to be more essentail for the
$B\to \pi\pi$ decays.

\bigskip
I thank members in the PQCD working group for useful discussions. The work
was supported in part by the National Science Council of R.O.C. under the
Grant No. NSC-91-2112-M-001-053, by the National Center for Theoretical
Science of R.O.C., and by Theory Group of KEK, Japan.

\end{document}